\documentclass[preprint]{aastex}

\usepackage{amsmath}
\usepackage{amssymb}
\usepackage{graphicx}
\usepackage{wasysym}
\usepackage{multirow}
\usepackage[breaklinks,colorlinks,citecolor=blue]{hyperref}

\slugcomment{Submitted to ApJ Letters}

\shorttitle{Protostellar Population in the CMZ}
\shortauthors{Lu et al.}

\begin{document}

\title{Deeply Embedded Protostellar Population in the 20 km\,s$^{-1}$ Cloud of the Central Molecular Zone}
	
\author{Xing Lu\altaffilmark{1,2,3,4}, Qizhou Zhang\altaffilmark{2}, Jens Kauffmann\altaffilmark{5}, Thushara Pillai\altaffilmark{5}, Steven N.~Longmore\altaffilmark{6}, J.~M.~Diederik Kruijssen\altaffilmark{7}, Cara Battersby\altaffilmark{2}, and Qiusheng Gu\altaffilmark{1,3,4}}
\affil{$^1$ School of Astronomy and Space Science, Nanjing University, Nanjing 210093, P.~R.~China; \email{xinglv.nju@gmail.com}}
\affil{$^2$ Harvard-Smithsonian Center for Astrophysics, 60 Garden Street, Cambridge, MA 02138, USA}
\affil{$^3$ Key Laboratory of Modern Astronomy and Astrophysics, Nanjing University, Nanjing 210093, P.~R.~China}
\affil{$^4$ Collaborative Innovation Center of Modern Astronomy and Space Exploration, Nanjing 210093, P.~R.~China}
\affil{$^5$ Max Planck Institut f\"{u}r Radioastronomie, Auf dem H\"{u}gel 69, D-53121 Bonn, Germany}
\affil{$^6$ Astrophysics Research Institute, Liverpool John Moores University, 146 Brownlow Hill, Liverpool L3 5RF, UK}
\affil{$^7$ Zentrum f\"{u}r Astronomie der Universit\"{a}t Heidelberg, Astronomisches Rechen-Institut, M\"{o}nchhofstra\ss e 12-14, 69120 Heidelberg, Germany}

\begin{abstract}
We report the discovery of a population of deeply embedded protostellar candidates in the 20 km\,s$^{-1}$ cloud, one of the massive molecular clouds in the Central Molecular Zone (CMZ) of the Milky Way, using interferometric submillimeter continuum and H$_2$O maser observations. The submillimeter continuum emission shows five 1-pc scale clumps, each of which further fragments into several 0.1-pc scale cores. We identify 17 dense cores, among which 12 are gravitationally bound. Among the 18 H$_2$O masers detected, 13 coincide with the cores and probably trace outflows emanating from the protostars. There are also 5 gravitationally bound dense cores without H$_2$O maser detection. In total the 13 masers and 5 cores may represent 18 protostars with spectral types later than B1 or potential growing more massive stars at earlier evolutionary stage, given the non-detection in the centimeter radio continuum. In combination with previous studies of CH$_3$OH masers, we conclude that the star formation in this cloud is at an early evolutionary phase, before the presence of any significant ionizing or heating sources. Our findings indicate that star formation in this cloud may be triggered by a tidal compression as it approaches pericenter, similar to the case of G0.253+0.016 but with a higher star formation rate, and demonstrate that high angular resolution, high sensitivity maser and submillimeter observations are a promising technique to unveil deeply embedded star formation in the CMZ.
\end{abstract}

\keywords{Galaxy: center --- ISM: clouds --- stars: formation}

\section{INTRODUCTION}\label{sec:intro}

The inner 500-pc region of the Galaxy, known as the Central Molecular Zone (CMZ), is rich in dense molecular gas, but the current star formation is unusually inactive. Other than the few star forming regions such as Sgr~B2 \citep{goldsmith1990,qin2008}, Sgr~A complex \citep{ekers1983,yusefzadeh2010}, and Sgr~C \citep{kendrew2013}, most CMZ clouds appear to be inactive in star formation \citep[e.g., G0.253+0.016;][]{lis1994,longmore2012,kauffmann2013a,johnston2014,mills2015,rathborne2015}. This inactivity is in contrast to the general star formation relation that has been established for the Galactic disk clouds as well as external galaxies, which presents a good correlation between the amount of dense molecular gas and the star formation rate. The overall star formation rate in the CMZ is an order of magnitude lower than the prediction of this correlation \citep{longmore2013b}.

Recent theoretical works by \citet{kruijssen2014} and \citet{KK2015} point out that star formation in the CMZ could be regulated by the strength of turbulence: the inflowing gas through the Galactic bar drives strong turbulence, resulting in episodes of low star formation activity; when the turbulence finally dissipates, a burst of star formation takes place. Such dissipation of the turbulence could be induced by compressive tides during a close passage to the bottom of the gravitational potential near Sgr~A* \citep{longmore2013a}. In the simulations of \citet{kruijssen2015}, the massive molecular clouds in the CMZ are found to be in several streams of open trajectory centered at Sgr~A* with a radius of $\sim$100~pc. The 20 km\,s$^{-1}$ cloud, a massive \citep[$\gtrsim$1.6$\times$10$^5$~$M_\odot$;][]{bally2010} molecular cloud named after its radial velocity, appears to be close to pericenter passage with Sgr~A*, therefore could be in the process of turbulence dissipation. In this scenario, we expect to observe increasing dense gas fraction and signs of early star formation in the 20 km\,s$^{-1}$ cloud.

Previous interferometric observations have found one \ion{H}{2} region \citep{downes1979}, and several H$_2$O masers \citep{sjouwerman2002,caswell2011} in this cloud. However, the maser observations were sensitivity-limited (usually $\gtrsim$30~mJy\,beam$^{-1}$ per 0.2~km\,s$^{-1}$ channel), hence could miss faint sources that trace star formation of lower masses or at early evolutionary stages. Here we use interferometric submillimeter observations to trace dense cores, and interferometric H$_2$O maser observations with a sensitivity of $\sim$5 times higher than previous studies to trace embedded protostars. Throughout the paper, we adopt a distance to the Galactic Center of 8.4~kpc \citep{reid2009}.

\section{OBSERVATIONS AND DATA REDUCTION}\label{sec:obs}

\addtocounter{footnote}{7}

\subsection{SMA Observations}\label{subsec:smaobs}

In 2013, we observed a mosaic of eight positions in the 20 km\,s$^{-1}$ cloud with the Submillimeter Array\footnote{The Submillimeter Array is a joint project between the Smithsonian Astrophysical Observatory and the Academia Sinica Institute of Astronomy and Astrophysics and is funded by the Smithsonian Institution and the Academia Sinina.} (SMA) in its compact and sub-compact configurations at 230~GHz band. The primary beam size is $\sim$56\arcsec{}. All observations share the same correlator setup: the rest frequencies of 216.9--220.9~GHz are covered in one sideband, and 228.9--232.9~GHz are covered in the other sideband, with a uniform channel width of 0.812~MHz, equivalent to 1.1~km\,s$^{-1}$ at 230~GHz. Observation details are listed in \autoref{tab:obs}.

The visibility data were calibrated using \mbox{MIR}\footnote{\url{https://www.cfa.harvard.edu/~cqi/mircook.html}} and inspected and imaged using \mbox{MIRIAD} \citep{sault1995} and \mbox{CASA} \citep{mcmullin2007}. Continuum was extracted from line free channels and imaged using data from both sidebands. Spectral lines were split from the continuum-subtracted visibility data and were imaged separately. We used a robust parameter of 0.5 when CLEANing images. The resulting continuum image has a clean beam of 4.9\arcsec{}$\times$2.8\arcsec{} (equivalent to 0.20~pc$\times$0.11~pc) with a position angle of 5.2\arcdeg{} and an rms of 3~mJy\,beam$^{-1}$. Typical rms of spectral lines images is $\sim$0.13 Jy\,beam$^{-1}$ per 1.1~km\,s$^{-1}$ channel.

\subsection{VLA Observations}\label{subsec:vlaobs}
In 2013 May, we observed a mosaic of three positions in this cloud with the National Radio Astronomy Observatory (NRAO)\footnote{The National Radio Astronomy Observatory is a facility of the National Science Foundation operated under cooperative agreement by Associated Universities, Inc.} Karl G.\ Jansky Very Large Array (VLA) in the DnC configuration at K band, with a primary beam size of $\sim$2\arcmin{}. The WIDAR correlator was configured to cover the H$_2$O maser at 22.2~GHz using a 16~MHz subband in dual polarizations, as well as 1.3~cm continuum using eight 128~MHz subbands in full polarizations. For the H$_2$O maser, the subband was split into 1024 channels with a channel width of 15.6~kHz, equivalent to 0.2~km\,s$^{-1}$. Observation details are listed in \autoref{tab:obs}.

The visibility data were calibrated and imaged using \mbox{CASA}. Continuum was extracted from line free channels of the 128~MHz subbands, leading to a total bandwidth of 0.9~GHz. The robust parameter we used in CLEAN is 0.5. For the H$_2$O maser image, the resulting clean beam is 3.5\arcsec{}$\times$2.4\arcsec{} (equivalent to 0.14~pc$\times$0.10~pc) with a position angle of 5.7\arcdeg{}. The rms in each 0.2~km\,s$^{-1}$ channel is 6~mJy\,beam$^{-1}$, but can be significantly larger in channels where signals are dynamic range limited.

\section{RESULTS}\label{sec:results}

\subsection{SMA Dense Cores}\label{subsec:cores}

\begin{figure}
\centering
\includegraphics[width=1.0\textwidth]{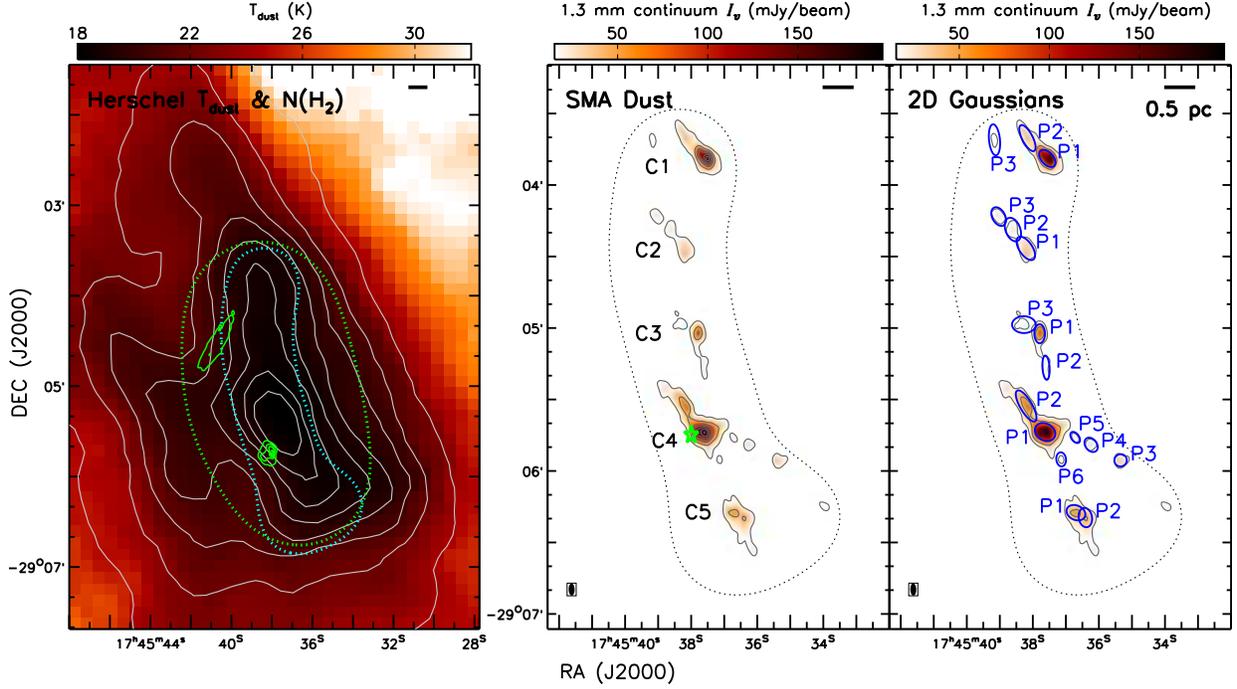}
\caption{Left: Herschel dust temperature is shown in the background image. The white contours present the Herschel H$_2$ column density, between 5$\times$10$^{22}$~cm$^{-2}$ and 4$\times$10$^{23}$~cm$^{-2}$ in step of 5$\times$10$^{22}$~cm$^{-2}$. The green contours present the VLA 1.3~cm continuum emission, between 2~mJy\,beam$^{-1}$ and 18~mJy\,beam$^{-1}$ in step of 4~mJy\,beam$^{-1}$. The cyan and green dotted loops show the FHWMs of the SMA and VLA primary beam responses, respectively. Middle: both contours and background image show the SMA 1.3~mm continuum emission. The contours are between 5$\sigma$ and 65$\sigma$ levels in step of 10$\sigma$, where 1$\sigma$=3~mJy\,beam$^{-1}$. The five clumps are labeled. The dotted loop shows the FWHM of the SMA primary beam response. The synthesized beam of the SMA is shown in the lower left corner. The \ion{H}{2} region is marked by a green star. Right: same as the middle panel, but only the 5$\sigma$ and 15$\sigma$ contours are plotted. The ellipses are the results of 2D Gaussian fittings.}
\label{fig:smacores}
\end{figure}

The SMA 1.3~mm continuum emission revealed five clumps of 1-pc scales in the 20 km\,s$^{-1}$ cloud, labelled as C1--C5 in \autoref{fig:smacores}. In the projected plane of the sky they are equally spaced by $\sim$1~pc and aligned along the densest part of the cloud. All the clumps appear to have substructures, or cores, among which C4 is the most prominent one which presents at least 6 cores.

After a visual inspection of dust peaks with fluxes above 5$\sigma$ levels, we identified 17 cores. Then we fitted 2D Gaussians to obtain their positions, deconvolved sizes, and primary-beam corrected continuum fluxes. We assumed a gas-to-dust mass ratio of 100, a dust emissivity index $\beta=1.5$, and applied the dust temperature $T_\text{dust}=19~\text{K}$ (see \autoref{fig:smacores}) and a mean continuum frequency of 225~GHz, then the core masses were calculated following \citet{beuther2005}. The results are listed in \autoref{tab:cores}.

\subsection{VLA H$_2$O Masers}\label{subsec:h2omasers}
\begin{figure}
\centering
\includegraphics[width=1.0\textwidth]{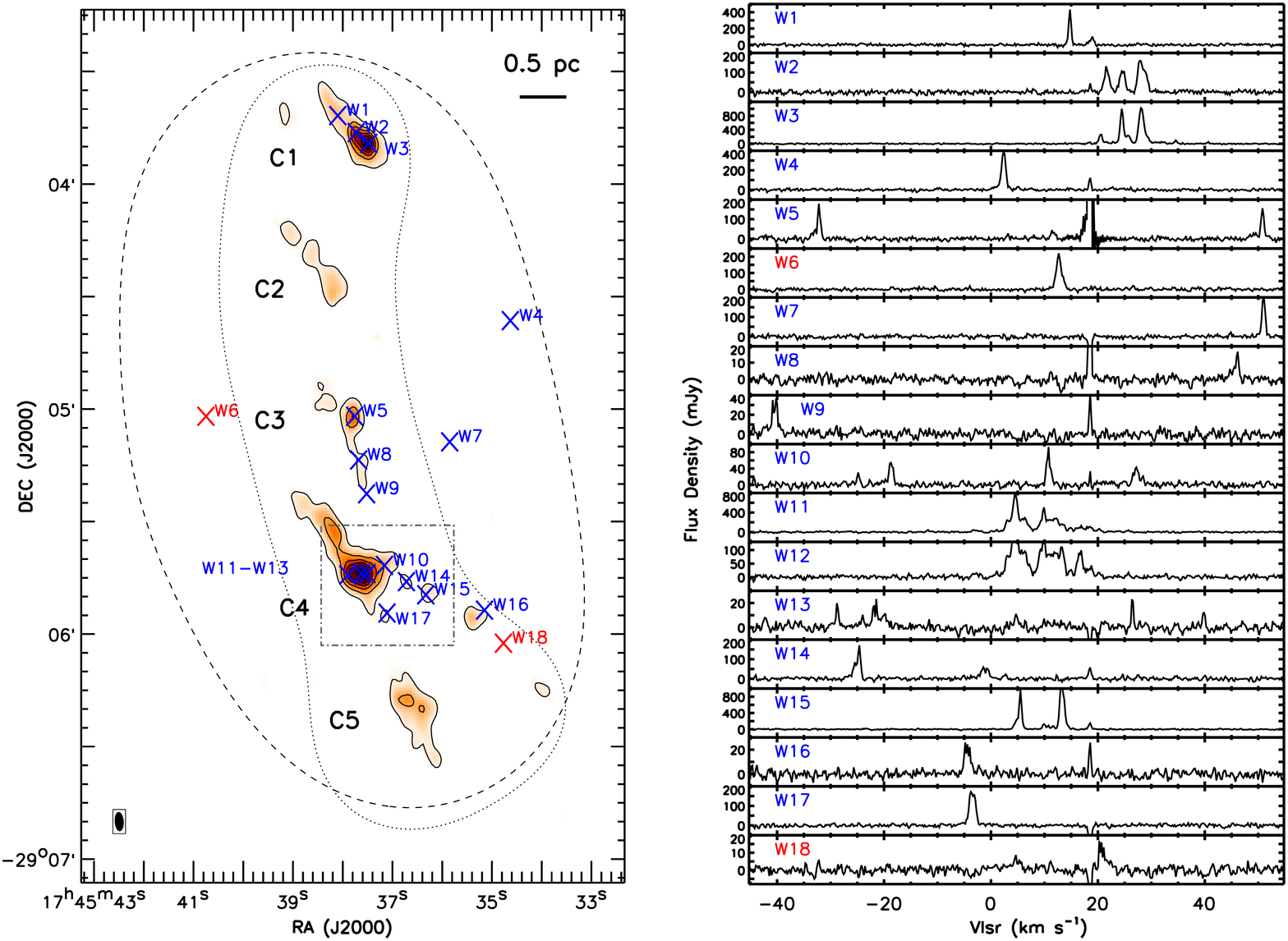}
\caption{Left: the crosses show the H$_2$O masers. The two red crosses (W6, W18) are known OH/IR stars. The blue crosses show H$_2$O masers without known OH/IR star counterparts. Both the contours and the background image show the SMA 1.3~mm continuum emission. The loops are the same as in \autoref{fig:smacores}. Right: the spectra of the 18 H$_2$O masers. The features at $\sim$18~km\,s$^{-1}$, either in emission or in absorption, are from sidelobes of the strongest maser, W5.}
\label{fig:masers}
\end{figure}

The VLA observations revealed 18 H$_2$O masers in this cloud, marked by crosses and labeled as W1--W18 in \autoref{fig:masers}. 15 out of 18 are within the SMA field, among which 13 spatially coincide with dust emission peaks above 5$\sigma$ levels. The velocities of these 13 masers are all within $\pm$20~km\,s$^{-1}$ with respect to the core velocities. Properties of these H$_2$O masers are summarized in \autoref{tab:masers} and their spectra are shown in \autoref{fig:masers}.

Among the previous H$_2$O maser surveys toward the CMZ, \citet{walsh2011} did not find any masers in this cloud using the Mopra telescope at a sensitivity of 1--2~Jy, while \citet{caswell2011} detected three masers, using ATCA at a sensitivity of $\lesssim$0.1~Jy: two of them are consistent with W13 and W15, respectively, within a position uncertainty of 2\arcsec{}; the other one is in C1, offset from the masers we detected by $\sim$3\arcsec{}. In addition, \citet{sjouwerman2002} serendipitously detected 10 H$_2$O masers in this cloud while studying OH/IR stars, all of which they concluded to be connected to star formation given the non-detection of OH/IR stars. One of them is consistent with W5, while the other 9 are scattered in C4: one is offset from any masers we detected, 8 are likely consistent with W10, W11--W13, and W15. 

\section{DISCUSSIONS}\label{sec:discussions}

\subsection{The Gravitational Equilibrium of the Dense Cores}\label{subsec:virial}

We analyze the virial status of the cores. The virial parameter is defined as $\alpha=5\sigma_v^2R/(GM_\mathrm{core})$ \citep{bertoldi1992,kauffmann2013b}, where $\sigma_v=\text{FWHM}/\sqrt{8\ln2}$ is the one-dimensional velocity dispersion. For a self-gravitating, non-magnetized core, the critical virial parameter is 2, above which the core is unbound and may expand, while below which it is bound and may collapse.

The N$_2$H$^+$ line has a critical density of $\gtrsim$10$^6$~cm$^{-3}$, and is superior for tracing dense gas than the spectral lines in our SMA data such as C$^{18}$O or H$_2$CO. To derive line widths of the cores, we obtain the SMA N$_2$H$^+$ 3--2 data (Kauffmann et al.\ in prep.), then fit Gaussians to the mean N$_2$H$^+$ spectra of the cores. For the 4 cores without N$_2$H$^+$ detections, we fit their mean C$^{18}$O or H$_2$CO spectra instead. Two cores, C3-P2 and C3-P3, do not present any of the three dense gas tracers, thus are excluded in the analysis. The results are listed in \autoref{tab:cores}.

Among the 15 cores included in the analysis, 12 have $\alpha<2$. The two most massive cores, C1-P1 and C4-P1, have virial parameters as low as $\sim$0.2. Three cores have $\alpha>2$, including C4-P4 which associates with an H$_2$O maser.

The masses themselves have large uncertainties: the errors in the core masses are a factor of 1.2--1.8, while the virial masses are sensitive to the selection of line widths and the errors are a factor of 1.7--2, according to the simulations of  \citet{battersby2010} that take errors in all variables of the mass estimates into account. Besides, the magnetic field might be dynamically important on pc scales in the CMZ clouds, as suggested by the ordered magnetic vectors in G0.253+0.016 \citep{pillai2015} which derive a magnetic flux density of $\sim$5~mG from the Chandrasekhar-Fermi method. Within these uncertainties, the 12 cores with $\alpha<2$ are gravitationally bound.

\subsection{The Nature of the H$_2$O Masers}\label{subsec:natureofmaser}

H$_2$O masers in star-forming regions are usually excited in shocked ambient gas \citep{elitzur1989}, therefore are used to trace protostellar outflows. However, they can also be excited in the envelope of evolved stars \citep{sjouwerman1996} or excited by shocks created in large-scale dynamic processes \citep[e.g., cloud collisions,][]{tarter1986}. We need to exclude these scenarios before using the H$_2$O masers as star formation indicators.

First, we compare the coordinates and velocities of the H$_2$O masers with the evolved star catalogues in \citet{lindqvist1992}, \citet{sevenster1997}, and \citet{sjouwerman1998,sjouwerman2002} which used OH or SiO masers as tracers. Two H$_2$O masers (W6, W18) are consistent with evolved stars (red crosses in \autoref{fig:masers}). The other 16 H$_2$O masers do not have known evolved star counterparts.

Second, we compare with the class I CH$_3$OH masers in \citet{yusefzadeh2013}, which are collisionally pumped and trace large-scale dynamics. None of the H$_2$O masers spatially coincide with the class I CH$_3$OH masers. Most of the H$_2$O masers coincide with the dense cores, instead of uniformly scatter like the class I CH$_3$OH masers. Their velocities are usually offset by $\pm$20~km\,s$^{-1}$ from the core velocities, which is easily seen in masers tracing outflows, instead of all showing the same value at the presumable shock interaction velocity. All these facts suggest that the H$_2$O masers are unlikely connected to large-scale shocks.

Therefore, the 16 H$_2$O masers without evolved star counterparts, in particular, the 13 masers coincident with the dense cores, are likely associated with outflows. Indeed, we find signatures of outflows associated with these cores traced by SiO, SO, and HNCO lines in the SMA images, which we will present in a forthcoming paper. A correlation between luminosities of H$_2$O masers and protostars \citep[e.g.,][]{palla1993} suggests that the more luminous masers ($\gtrsim$10$^{-6}$~$L_\odot$, see \autoref{tab:masers}) may correspond to early B-type stars of $\gtrsim$10$^3$~$L_\odot$, while the characteristic luminosity of $\sim$10$^{-7}$~$L_\odot$ for most masers is usually found in low- or intermediate-mass protostars \citep[e.g.,][]{furuya2001}.

\subsection{Comparison with Other Star Formation Tracers}\label{subsec:othersftracers}

The free-free emission from \ion{H}{2} regions is a reliable star formation tracer. However, only one \ion{H}{2} region has been found in this  cloud \citep{downes1979}, which is verified by our VLA 1.3~cm continuum observation (\autoref{fig:smacores}). Assuming an electron temperature of 10$^4$~K, with a continuum flux of 0.12~Jy, its ionizing photon rate is 9$\times$10$^{47}$~s$^{-1}$ \citep{mezger1967}, corresponding to an O9 star of 19~$M_\odot$ \citep{mottram2011}. The non-detection of any other \ion{H}{2} regions at a sensitivity of 100~$\mu$Jy may suggest no protostars earlier than B1 above $\sim$10~$M_\odot$, or simply an early evolutionary phase without any visible ionizing sources.

\citet{yusefzadeh2009} identified YSOs in the CMZ using the \textit{Spizter} 8~\micron{} and 24~\micron{} emission, and found three in the vicinity of this cloud, but none of them are within the cloud itself. The only visible infrared point source within the cloud in the \textit{Spitzer} mid-infrared images is the \ion{H}{2} region. Therefore, the deeply embedded protostars are not observable in the infrared bands, probably due to the large column density ($\gtrsim$10$^{23}$~cm$^{-2}$) which obscures the embedded sources.

CH$_3$OH masers have been found in star-forming regions in the Galactic disk, and are classified as class I or class II, depending on whether they are collisionally or radiatively pumped \citep{menten1991a,fontani2010}. Recent class I CH$_3$OH maser observations toward the CMZ \citep{yusefzadeh2013,mills2015} suggested that these masers may not trace star formation, but may have cosmic ray or cloud-scale dynamic origins. On the other hand, radiatively pumped class II CH$_3$OH masers have been reliable tracers of high-mass star formation \citep{urquhart2015}. The methanol multi-beam survey \citep[MMB,][]{caswell2010} did not detect any class II (6.7~GHz) CH$_3$OH masers at a sensitivity of 0.17~Jy in this cloud. By contrast, four H$_2$O masers in our  results (W3, W5, W11, W15) are detectable at 3$\sigma$ levels with the same sensitivity. 

\subsection{Implications for Star Formation in the 20 km\,s$^{-1}$ Cloud}\label{subsec:sfhistory}

As discussed above, we find 13 H$_2$O masers associated with the dense cores and probably trace protostellar outflows, and 5 dense cores without H$_2$O maser detection but gravitationally bound. In total, they may represent 18 protostellar candidates. This number should be a lower limit because the dense cores could harbor multiple protostars. By contrast, previous observations found $\sim$10 H$_2$O masers most of which are concentrated in C4, corresponding to 6 protostellar candidates in our sample. In the following we discuss the implication of our results for the evolutionary phase and star formation activity of the 20 km\,s$^{-1}$ cloud.

The evolutionary phases of star formation traced by H$_2$O masers is under debate \citep[e.g.,][]{szymczak2005,breen2010}. Recent follow-up studies of the unbiased MMB survey suggest that while not as well-defined as the other masers, H$_2$O masers in general trace the protostellar phase when outflows emerge \citep{breen2014,titmarsh2014}. On the other hand, these studies seem to converge to the conclusion that class II CH$_3$OH masers trace a well-defined evolutionary phase (e.g., after protostars start to heat ambient gas and before the UC \ion{H}{2} region phase). For the 20 km\,s$^{-1}$ cloud, the large population of H$_2$O masers and the virtually non-detection of class II CH$_3$OH masers so far, combining with the non-detection of any prominent UC \ion{H}{2} regions, are likely indicative of two alternative scenarios: an early evolutionary phase when high-mass protostars have not started to heat or ionize the ambient gas, but have developed outflows, similar to the situation in infrared dark clouds (IRDCs); or a cluster of low- or intermediate-mass protostars, whose outflows power the H$_2$O masers but whose radiation is not enough to create class~II CH$_3$OH masers or visible \ion{H}{2} regions. The former may be preferable because the more luminous masers may trace high-mass protostars (see Section~\ref{subsec:natureofmaser}).

No matter which scenario it is, the 18 protostellar candidates traced by H$_2$O masers and gravitationally bound cores suggest more active star formation than that traced by free-free or infrared emission (one \ion{H}{2} region). Sensitive maser and submillimeter observations could be a promising method to trace incipient star formation in other massive clouds in the CMZ.

In the orbital models of \citet{kruijssen2015}, G0.253+0.016 is also close to pericenter passage with Sgr~A*, but in a different gas stream. G0.253+0.016 has a similar mass as the 20 km \,s$^{-1}$ cloud \citep[$\sim$1.3$\times$10$^5$~$M_\odot$;][]{longmore2012}, but only has one known H$_2$O maser \citep{lis1994} and one gravitationally bound dense core associated with the maser \citep{kauffmann2013a,rathborne2015}. The progression of star formation activity from G0.253+0.016 to Sgr~B2 supports the tidal compression and triggered collapse model proposed by \citet{longmore2013a}. The 20 km\,s$^{-1}$ cloud could be a second case of such triggering. Based on the number of masers and the amount of dense gas contained in cores, the current star formation rate in the 20 km\,s$^{-1}$ cloud is likely higher than in G0.253+0.016. The difference might be due to variations in the local environment in the streams.

\section{CONCLUSIONS}\label{sec:conclusions}
We have found 18 protostellar candidates traced by dust emission and H$_2$O masers in the 20 km\,s$^{-1}$ cloud, most of which have been missed by previous infrared or radio continuum studies. Systematic studies of other massive clouds in the CMZ, using high resolution, sensitive maser and submillimeter observations, will be helpful to establish their star formation status. One such example is the SMA Legacy Survey of the CMZ\footnote{\url{https://www.cfa.harvard.edu/sma/largeScale/CMZ/}} (PIs: C.~Battersby \& E.~Keto) that will cover all regions in the CMZ above a column density threshold of 10$^{23}$~cm$^{-2}$ using the same setups as our SMA observations.

A virial analysis shows that among the 17 dense cores traced by dust emission, 12 cores are gravitationally bound. The 13 H$_2$O masers associated with the dense cores likely trace protostellar outflows. The star formation in the 20 km\,s$^{-1}$ cloud appears to be in an early evolutionary phase, before significant heating or ionization of ambient gas. This cloud may follow the tidal compression and triggered collapse model of \citet{longmore2013a} and \citet{kruijssen2015}, similar to the dust ridge clouds. However, its star formation rate is higher than in G0.253+0.016, which likely reflects local differences in initial conditions.

\acknowledgments
We thank the anonymous referee for constructive comments, and Elisabeth Mills and Hauyu Baobab Liu for helpful discussion. X.L.\ acknowledges the support of a Smithsonian Predoctoral Fellowship and the program A for outstanding PhD candidate of Nanjing University. T.P. acknowledges financial support from the \emph{Deut\-sche For\-schungs\-ge\-mein\-schaft, DFG\/} via the SPP (priority program) 1573 (``Physics of the ISM"). J.M.D.K.\ is funded by a Gliese Fellowship. This work is supported by the SI CGPS grant on Star Formation in the Central Molecular Zone of the Milky Way, the National Natural Science Foundation of China (grants 11328301, 11273015 and 11133001), and the National Basic Research Program (973 program No.~2013CB834905).

\clearpage

\begin{deluxetable}{cccccccccccc}
\tabletypesize{\scriptsize}
\rotate
\tablecaption{Summary of the observations. \label{tab:obs}}
\tablewidth{0pt}
\tablehead{
\multirow{2}{*}{Telescope} & \multirow{2}{*}{PI} & \multirow{2}{*}{Project ID} & \multirow{2}{*}{Lines} & \multirow{2}{*}{Date} & \multirow{2}{*}{$\tau_\text{225~GHz}$} & $T_\text{sys}$ & \multirow{2}{*}{Pointing\tablenotemark{a}} & \multicolumn{3}{c}{Calibrators} & \multirow{2}{*}{Note} \\
\cline{9-11} & & & & & & (K) & & Bandpass & Flux & Gain\tablenotemark{b} &
}
\startdata
SMA 	& X.~Lu		& 2013A-S049 & Many &2013 Jul 24& 0.25 & 100--400 &  S1--S8 & 3C84 & Neptune & Q1, Q2 & 6 antennas  \\
Compact	&			&		         &	     &2013 Aug 03& 0.10 & 100--240 & S1--S8 & 3C84 & Neptune & Q1, Q2  & 5 antennas\\
 		&			&		         &	     &2013 Aug 09& 0.20 & 100--240 &S1--S8 & 1924$-$292 & Neptune & Q1, Q2 & 5 antennas \\

SMA & \multirow{2}{*}{Q.~Zhang} 	& \multirow{2}{*}{2012B-S097} & \multirow{2}{*}{Many} &\multirow{2}{*}{2013 May 21}& \multirow{2}{*}{0.17} & \multirow{2}{*}{120--180} & \multirow{2}{*}{S1--S8} & \multirow{2}{*}{3C279} & Neptune, & \multirow{2}{*}{Q1, Q2} & \multirow{2}{*}{5 antennas} \\
Subcompact & & & & & & & & & Titan & & \\

\multirow{3}{*}{VLA DnC} & \multirow{3}{*}{Q.~Zhang}  & \multirow{3}{*}{13A-307} & NH$_3$\,(1,1)--(5,5), &\multirow{3}{*}{2013 May 12}& \multirow{3}{*}{\nodata} & \multirow{3}{*}{\nodata} &  \multirow{3}{*}{V1--V3} & \multirow{3}{*}{3C279} & \multirow{3}{*}{3C286} & \multirow{3}{*}{Q2} & \multirow{3}{*}{\nodata} \\
 & & & H$_2$O maser,  & & & & & & & & \\
 & & & 1.3~cm continuum & & & & & & & &
\enddata
\tablenotetext{a}{Coordinates of pointing centers: S1: (17:45:38.35, $-$29:03:49.90); S2: (17:45:38.77, $-$29:04:18.60); S3: (17:45:38.64, $-$29:04:46.20); S4: (17:45:38.10, $-$29:05:13.90); S5: (17:45:37.55, $-$29:05:40.40); S6: (17:45:36.80, $-$29:06:07.60); S7: (17:45:35.00, $-$29:06:19.10); S8: (17:45:36.79, $-$29:06:31.10); V1: (17:45:38.60, $-$29:04:09.50); V2: (17:45:38.00, $-$29:05:08.80); V3: (17:45:36.60, $-$29:06:05.00).}
\tablenotetext{b}{Gain calibrators: Q1: 1733$-$130; Q2: 1744$-$312.}
\end{deluxetable}

\begin{deluxetable}{cccrrccc}
\tabletypesize{\scriptsize}
\tablecaption{Properties of the dense cores. \label{tab:cores}}
\tablewidth{0pt}
\tablehead{
\multirow{2}{*}{Core ID} & R.A. \& Decl. & Maj. $\times$ Min.\tablenotemark{a} & PA\tablenotemark{a} & Flux\tablenotemark{b} & FWHM\tablenotemark{c} & $M_\text{core}$ & $\alpha$ \\
 & (J2000) & (\arcsec{}$\times$\arcsec{}) & (\arcdeg{}) & (mJy) & (km\,s$^{-1}$) & ($M_\odot$) &
 }
\startdata
C1-P1 & 17:45:37.58, -29:03:48.83 & $7.22\times3.16$ & 48.8 & 648.1 & 3.0 (N) & 1309 & 0.17 \\
C1-P2 & 17:45:38.18, -29:03:40.31 & $11.1\times2.75$ & 28.4 & 191.7 & 2.9 (N) & 387 & 0.62 \\
C1-P3 & 17:45:39.17, -29:03:41.03 & $12.2\times3.32$ & 5.25 & 112.7 & 4.2 (N) & 228 & 2.32 \\

C2-P1 & 17:45:38.23, -29:04:26.60 & $10.1\times4.10$ & 36.0 & 189.0 & 5.8 (N) & 382 & 2.55 \\
C2-P2 & 17:45:38.62, -29:04:18.69 & $9.10\times5.02$ & 22.7 & 118.5 & 3.2 (N) & 239 & 1.45 \\
C2-P3 & 17:45:39.04, -29:04:13.24 & $6.98\times3.96$ & 34.7 & 87.8   & 2.5 (N) & 177 & 1.02 \\

C3-P1 & 17:45:37.81, -29:05:02.41 & $6.38\times3.12$ & 178.0 & 208.1& 4.3 (H) & 420 & 0.92 \\
C3-P2 & 17:45:37.62, -29:05:16.65 & $8.99\times0.49$ &  3.1     & 63.0  & \nodata & 127 & \nodata \\
C3-P3 & 17:45:38.28, -29:04:58.59 & $8.96\times5.11$ & 90.8   & 112.7 & \nodata & 228 & \nodata \\

C4-P1 & 17:45:37.64, -29:05:43.65 & $7.85\times5.35$ & 68.0 & 935.4 & 3.6 (N) & 1889 & 0.22 \\ 
C4-P2 & 17:45:38.23, -29:05:32.72 & $14.0\times3.51$ & 30.5 & 393.2 & 3.6 (C) & 794 & 0.56 \\
C4-P3 & 17:45:35.36, -29:05:55.53 & $4.07\times1.70$ & 99.0 & 104.1 & 3.8 (C) & 210 & 0.87 \\
C4-P4 & 17:45:36.25, -29:05:49.03 & $5.00\times2.70$ & 56.0 & 76.4   & 5.0 (N) & 154 & 2.73 \\
C4-P5 & 17:45:36.74, -29:05:45.93 & $<$5.23~$\times$~$<$3.07 & \nodata & 35.3  & 2.4 (N) & 71 & 1.82 \\
C4-P6 & 17:45:37.16, -29:05:55.13 & $3.41\times2.63$ &  6.2 & 37.9 & 2.3 (C) & 76 & 1.19 \\

C5-P1 & 17:45:36.71, -29:06:17.50 & $7.12\times3.93$ & 73.4 & 189.7 & 2.7 (N) & 383 & 0.54 \\
C5-P2 & 17:45:36.43, -29:06:19.55 & $6.57\times4.45$ & 13.1 & 151.2 & 2.3 (N) & 305 & 0.54
\enddata
\tablenotetext{a}{Major and minor FWHMs and position angles of the cores are deconvolved from the beam.}
\tablenotetext{b}{Fluxes are corrected for primary-beam response.}
\tablenotetext{c}{\mbox{Letters} in parentheses indicate the lines used to estimate FWHM: N - N$_2$H$^+$; C - C$^{18}$O; H - H$_2$CO.}
\end{deluxetable}

\begin{deluxetable}{ccccccc}
\tabletypesize{\scriptsize}
\tablecaption{Properties of the H$_2$O masers. \label{tab:masers}}
\tablewidth{0pt}
\tablehead{
\multirow{2}{*}{Maser ID} & R.A. \& Decl. & $v_\text{peak}$\tablenotemark{a} & $F_\text{peak}$\tablenotemark{a} & $F_\text{integrated}$\tablenotemark{b} & $L_\text{H$_2$O}$ & \multirow{2}{*}{Dense Cores} \\
 & (J2000) & (km\,s$^{-1}$) & (mJy) & (mJy$\cdot$km\,s$^{-1}$) & (10$^{-7}$~$L_\odot$) & 
 }
\startdata
W1 & 17:45:38.10, -29:03:41.75 & 14.8, 19.0 & 431, 100 & 430 & 7.0 & C1-P2 \\
W2 & 17:45:37.73, -29:03:46.29 & 21.5, 27.8 & 131, 161 & 775 & 12.6 & C1-P1\\
W3 & 17:45:37.48, -29:03:49.15 & 24.5, 28.1 & 999, 1022 & 3797 & 61.8 & C1-P1\\
W4 & 17:45:34.63, -29:04:36.42 & 2.4 & 396 & 670 & 10.9 & \nodata \\
W5 & 17:45:37.76, -29:05:01.91 & -32.2, 18.6 & 180, 13180 & 6822 & 110.0 & C3-P1\\
W6 & 17:45:40.75, -29:05:01.93 & 12.7 & 225 & 357 & 5.8 & \nodata \\
W7 & 17:45:35.85, -29:05:08.75 & 51.0 & 212 & 186 & 3.0 & \nodata \\
W8 & 17:45:37.68, -29:05:13.57 & 46.2 & 17 & 18 & 0.3 & C3-P2 \\
W9 & 17:45:37.52, -29:05:22.59 & -40.8, -40.2 & 33, 35 & 46 & 0.7 & \nodata \\
W10 & 17:45:37.16, -29:05:41.70 & -18.7, 10.8 & 54, 93 & 196 & 3.2 & C4-P1\\
W11 & 17:45:37.62, -29:05:43.92 & 4.5, 9.9 & 783, 507 & 3258 & 53.0 & C4-P1 \\
W12 & 17:45:37.51, -29:05:43.89 & 13.1, 16.9 & 110, 99 & 264 & 4.3 & C4-P1 \\
W13 & 17:45:37.90, -29:05:44.24 & -21.5, 26.4 & 23, 23 & 55 & 0.9 & C4-P1 \\
W14 & 17:45:36.72, -29:05:46.02 & -25.5, -24.6 & 84, 171 & 246 & 4.0 & C4-P5 \\
W15 & 17:45:36.33, -29:05:49.52 & 5.5, 13.1 & 972, 1178 & 2377 & 38.7 & C4-P4 \\
W16 & 17:45:35.15, -29:05:53.62 & -4.8, -4.4 & 26, 24 & 48 & 0.8 & C4-P3\\
W17 & 17:45:37.11, -29:05:54.38 & -3.8, -3.1 & 171, 159 & 302 & 4.9 & C4-P6 \\
W18 & 17:45:34.77, -29:06:02.43 & 20.3, 20.7 & 17, 16 & 33 & 0.5 & \nodata
\enddata
\tablenotetext{a}{Peak fluxes are not corrected for primary-beam response. For masers with more than two velocity components, only the two strongest peaks are listed.}
\tablenotetext{a}{Integrated fluxes are corrected for primary-beam response.}
\end{deluxetable}

\clearpage

\clearpage

\end{document}